\documentclass[fleqn,twoside]{article}
\usepackage[headings]{espcrc2}

\usepackage{bm}
\readRCS
$Id: espcrc2.tex,v 1.2 2004/02/24 11:22:11 spepping Exp $
\usepackage{graphicx}
\usepackage[figuresright]{rotating}

\newcommand{\AmS}{{\protect\the\textfont2
  A\kern-.1667em\lower.5ex\hbox{M}\kern-.125emS}}
\newcommand{\be}{\begin{equation}}
\newcommand{\ee}{\end{equation}}
\newcommand{\bea}{\begin{eqnarray}}
\newcommand{\eea}{\end{eqnarray}}
\def\avk{\langle k_\perp ^2\rangle}
\def\avp{\langle p_\perp ^2\rangle}
\def\pp{p_\perp}

\def\T{_{_T}}
\def\C{_{_C}}

\def\avPT{\langle P_T^2\rangle}
\newcommand{\ua}{\uparrow}

\newcommand{\bfp}{\mbox{\boldmath $p$}}

\def\bpp{\bfp_\perp}
\newcommand{\pup}{p^\uparrow}
\newcommand{\qup}{q^\uparrow}

\title{Update on transversity and Collins functions from SIDIS and
  $e^+e^-$ data} 

\author{M.~Anselmino\address[to]{Dipartimento di Fisica Teorica, Universit\`a di Torino and \\
        INFN, Sezione di Torino, Via P. Giuria 1, I-10125 Torino,
        Italy},
        M. Boglione\addressmark[to],
        U. D'Alesio\address[unica]{Dipartimento di Fisica, Universit\`a di Cagliari,\\
        Cittadella Universitaria di Monserrato, I-09042 Monserrato (CA),
        Italy}\address[infnca]{INFN, Sezione di Cagliari, C.P. 170, I-09042 Monserrato (CA),
        Italy}\thanks{Talk delivered at the Ringberg Workshop ``New
        Trends in HERA Physics 2008'', Ringberg Castle, Tegernsee,
        Germany, October 5-10, 2008.},  
        A. Kotzinian\address{CEA-Saclay, IRFU/Service de Physique
        Nucl\'eaire, 91191 Gif-sur-Yvette, 
        France; \\ Yerevan Physics Institute, 375036 Yerevan, Armenia; JINR,
        141980 Dubna, Russia},
        F. Murgia\addressmark[infnca],
        A. Prokudin\addressmark[to],
        and S. Melis\addressmark[to]}

\runauthor{M. Anselmino et al.}

\begin{document}

\begin{abstract}

We present an update of a previous global analysis of the experimental
data on azimuthal asymmetries in semi-inclusive deep inelastic
scattering (SIDIS), from the HERMES and COMPASS Collaborations, and
in $e^+e^- \to h_1 h_2 X$ processes, from the Belle Collaboration.
Compared to the first extraction, a more precise determination of
the Collins fragmentation function and the transversity distribution
function for $u$ and $d$ quarks is obtained.
\vspace{-0.2pc}
\end{abstract}

\maketitle

\section{Introduction}

The study of the nucleon spin structure has recently made remarkable
progress. Our understanding of the longitudinal degrees of freedom,
concerning both the intrinsic motion and the spin content of partons
inside unpolarized and longitudinally polarized fast moving nucleons,
respectively encoded in the unpolarized quark distribution, $q(x)$,
and the helicity distribution, $\Delta q(x)$, is now quite accurate.

A full knowledge of the nucleon quark structure in the collinear,
$\bm{k}_\perp$ integrated, configuration cannot be reached without
information on the third twist-two parton distribution function: the
transversity distribution, $\Delta_T q(x)$ (also denoted as
$h_1(x)$). Despite its fundamental importance~\cite{Ralston:1979ys}
and the intense theoretical work of the last
decade~\cite{Barone:2001sp}, this function has only very recently
been accessed experimentally. The main difficulty in measuring
transversity is that, being a chiral-odd quantity, it decouples from
inclusive deep inelastic scattering (DIS), since perturbative QED
and QCD interactions cannot flip the chirality of quarks.

The only way to access this distribution is by coupling it to
another chiral-odd quantity. To such a purpose one can look for a
chiral-odd partner either in the initial or the final state. In the
first case the most promising approach, and the cleanest one from
the theoretical point of view, is the study of the double transverse
spin asymmetry, $A_{TT}$, in Drell-Yan processes. This measurement
is in principle feasible at RHIC, but the small $x$ region covered
at a c.m.~energy $\sqrt s = 200$ GeV and the fact that in $pp$
collisions one measures the product of two transversity
distributions, one for a quark and one for an antiquark, lead to
$A_{TT}$ values of the order of a few percents~\cite{Martin:1999mg}.
A much larger $A_{TT}$, around 20-40\%, could be observed in
Drell-Yan processes in $p \bar p$ interactions at $s\simeq 200$
GeV$^2$, as proposed by the PAX
Collaboration~\cite{Barone:2005pu,Anselmino:2004ki,Efremov:2004qs,Pasquini:2006iv}.
However, this requires the availability of polarized antiprotons,
which is an interesting, but formidable task in itself. Other double
transverse spin asymmetries for inclusive production of photons or
pions are strongly suppressed by the large gluon contribution in the
unpolarized cross sections~\cite{Mukherjee:2003pf,Mukherjee:2005rw}.

For the case of a chiral-odd partner in the final state, we mention
the spin transfer in $\pup p \to \Lambda^\uparrow X$ or $\ell \,
\pup \to \ell'\Lambda^\uparrow X$ processes, where the final hyperon
acts as a polarimeter. Here $\Delta_T q$ couples to another unknown
quantity: the fragmentation function for a transversely polarized
quark into a transversely polarized baryon.

Recently, a promising suggestion has been made and is going to be
pursued: the combined study of two-hadron inclusive production in
single polarized DIS and in electron-positron annihilation
measurements~\cite{Bacchetta:2008wb}. In this context the new
quantity which appears is the so-called dihadron fragmentation
function describing the hadronization of a quark in two
hadrons~\cite{Efremov:1992pe,Collins:1994ax,Artru:1995zu,Jaffe:1997hf,Bacchetta:2003vn}.

Meanwhile, the most accessible and fruitful channel is the azimuthal
asymmetry $A_{UT}^{\sin(\phi_h + \phi_S)}$ in SIDIS processes,
namely $\ell \, \pup \to \ell \, h \, X$, involving the convolution
of the transversity distribution with the Collins fragmentation
function~\cite{Collins:1992kk}: the spin and transverse momentum
dependent (TMD) function parameterizing a left-right asymmetry in
the fragmentation of a transversely polarized quark into an
unpolarized hadron.

This study has been and still is under active investigation by the
HERMES, COMPASS and JLab Collaborations.

A crucial breakthrough has been achieved thanks to the independent
measurement of the Collins function (or rather, of the convolution
of two Collins functions), in $e^+e^- \to h_1 h_2 \, X$ unpolarized
processes by the Belle Collaboration at KEK \cite{Abe:2005zx}. By
combining the SIDIS data from HERMES~\cite{Airapetian:2004tw} and
COMPASS~\cite{Ageev:2006da}, with the Belle data, a global fit
leading to the first extraction of the transversity distribution and
the Collins fragmentation functions was performed in
Ref.~\cite{Anselmino:2007fs}.

Recently, much higher statistics data on these spin azimuthal
asymmetries have become available: the HERMES Collaboration have
presented charged and neutral pion, as well as kaon azimuthal
asymmetries~\cite{Diefenthaler:2007rj}; the COMPASS Collaboration
have presented their measurements, still on a deuteron target, but
now for separate charged pion and kaon production~\cite{:2008dn};
the Belle Collaboration have issued new high-precision data of the
Collins asymmetry in $e^+e^-$ annihilation~\cite{Seidl:2008xc}.

Therefore, we reconsider here our previous analysis and study the
impact of these new data on the extraction of both the transversity
and the Collins functions.

\section{Formalism}

We recall here the main steps necessary to calculate the azimuthal
asymmetries in SIDIS and in $e^+e^-$ annihilation, addressing to
Ref.~\cite{Anselmino:2007fs} for details. This approach is based on
an extension of the ordinary collinear factorization theorems with
inclusion of a new class of spin and TMD
distributions~\cite{Mulders:1995dh,Boer:1997nt,Bacchetta:2006tn,D'Alesio2007jt}.
For such processes $k_\perp$ factorization has been
proven~\cite{Collins:1981uk,Ji:2004xq,Ji:2004wu} in the regime of
low observed transverse momenta (compared to the large scale of the
processes). Another important result, crucial for this analysis, is
the universality of the Collins function entering SIDIS and $e^+e^-$
processes, as discussed in Refs.~\cite{Metz:2002iz,Collins:2004nx}.
We will restrict ourselves to tree level expressions, as currently
done in phenomenological studies, neglecting the soft factor coming
from gluon resummation~\cite{Collins:1981uk,Ji:2004xq,Ji:2004wu}
responsible of potential Sudakov suppression~\cite{Boer:2001he}. We
will briefly comment on this in the sequel.

\subsection{SIDIS}

Let us consider the single spin asymmetry for the process $\ell
\, p^\uparrow\to\ell^\prime h  X$:
 \bea A_{UT} &=& \frac{\displaystyle
      d^6\sigma^{\ell p^\uparrow \to \ell^\prime h X} \,-\,
      d^6\sigma^{\ell p^\downarrow \to \ell^\prime h X}}
     {\displaystyle
      d^6\sigma^{\ell p^\uparrow \to \ell^\prime h X} \,+\,
      d^6\sigma^{\ell p^\downarrow \to \ell^\prime h X}} \nonumber\\
      & \equiv &
\frac{d\sigma^\uparrow - d\sigma^\downarrow}
     {d\sigma^\uparrow + d\sigma^\downarrow}\,,
\label{sidis-asym}
 \eea
 where $d^6\sigma^{\ell p^{\uparrow,\downarrow} \to \ell
h X} \equiv d\sigma^{\uparrow,\downarrow}$ is a shorthand notation
for $(d^6\sigma^{\ell p^{\uparrow,\downarrow} \to \ell h X })/(d x
\, dy \, dz \, d^2 \bm{P}_T \, d\phi_S)$ with $x,y,z$ the usual
SIDIS variables. The $\phi_S$ dependence originates from the cross
section dependence on the angle between the proton (transverse)
polarization vector and the leptonic plane. We adopt here the
standard SIDIS kinematics according to the ``Trento
Conventions"~\cite{Bacchetta:2004jz}, see
also~\cite{Anselmino:2007fs}. By considering the $\sin(\phi_h
+\phi_S)$ moment of $A_{UT}$, we are able to single out the effect
coming from the spin dependent part of the fragmentation function
 of a transversely polarized quark (embedded in the Collins function,
$\Delta^N\!D_{h/q^\uparrow}(z,p_\perp)$ or $H_1^{\perp
q}(z,\pp)$~\cite{Bacchetta:2004jz}) coupled to the TMD transversity
distribution ($\Delta_T q(x,k_\perp)$). More explicitly, we get
 \bea
&& \hspace*{-.5cm}A^{\sin (\phi_h + \phi_S)}_{UT}  \nonumber\\
&& = 2 \, \frac{\int d\phi_S \, d\phi_h \, [d\sigma^\uparrow -
d\sigma^\downarrow] \, \sin(\phi_h +\phi_S)} {\int d\phi_S \,
d\phi_h \, [d\sigma^\uparrow +
d\sigma^\downarrow]}\nonumber\\
 && \propto 
 \frac{\sum_q e_q^2\, \Delta_T q(x,k_\perp) \otimes
\Delta^N\!D_{h/q^\uparrow}(z,p_\perp)}{\sum_q e_q^2\,
f_{q/p}(x,k_\perp)\otimes D_{h/q}(z,p_\perp)}\,, \label{defsin-asym}
 \eea
where $\otimes$ stands for a convolution on the transverse momenta
(see Eq.~(4) of Ref.~\cite{Anselmino:2007fs} for full details and
related comments).

This analysis can be further simplified by working at ${\cal
O}(k_\perp /Q)$ and adopting a Gaussian and factorized
parameterization of TMDs. In particular for the unpolarized parton
distribution (PDF) and fragmentation (FF) functions we use:
 \bea
 f_{q/p}(x,k_\perp) & = &
 f_{q/p}(x)\;\frac{e^{-{k_\perp^2}/\avk}}{\pi\avk}\\
 D_{h/q}(z,\pp)& =& D_{h/q}(z)\;\frac{e^{-\pp^2/\avp}}{\pi\avp}\;,
 \eea
with $\langle k_\perp^2\rangle$ and $\langle p_\perp^2\rangle$ fixed
to the values found in Ref.~\cite{Anselmino:2005nn} by analyzing
unpolarized SIDIS:
 \be
 \langle k_\perp^2\rangle = 0.25  \; {\rm
GeV}^2\,, \;\; \langle p_\perp^2\rangle  = 0.20 \;{\rm GeV}^2 \,.
 \ee

Integrated parton distribution and fragmentation functions,
$f_{q/p}(x)$ and $D_{h/q}(z)$, are available in the literature; in
particular, we use the GRV98LO PDF set~\cite{Gluck:1998xa} and the
DSS fragmentation function set~\cite{deFlorian:2007aj}.

{F}or the transversity distribution, $\Delta_T q(x, k_\perp)$, and
the Collins FF, $\Delta^N\! D_{h/q^\uparrow}(z,\pp)$, we adopt the
following parameterizations~\cite{Anselmino:2007fs}:
 \bea
\Delta_T q(x, k_\perp) &=&\frac{1}{2} {\cal N}^{\T}_q(x)\,
[f_{q/p}(x)+\Delta q(x)] \nonumber\\
& & \times \frac{e^{-{k_\perp^2}/{\avk\T}}}{\pi \avk \T} \label{tr-funct}\\
 \Delta^N \! D_{h/q^\uparrow}(z,\pp) &=& 2 {\cal N}^{\C}_q(z)\,
D_{h/q}(z)\nonumber\\
&& \times h(\pp)\,\frac{e^{-\pp^2/{\avp}}}{\pi \avp}\,,
\label{coll-funct}
 \eea
 with
 \bea
 {\cal N}^{\T}_q(x)= N^{\T}_q
\,x^{\alpha} (1-x)^{\beta} \, \frac{(\alpha + \beta)^{(\alpha
+\beta)}} {\alpha^{\alpha} \beta^{\beta}}
\\
{\cal N}^{\C}_q(z)= N^{\C}_q \, z^{\gamma} (1-z)^{\delta} \,
\frac{(\gamma + \delta)^{(\gamma +\delta)}}
{\gamma^{\gamma} \delta^{\delta}}\\
h(\pp)=\sqrt{2e}\,\frac{p_\perp}{M_{h}}\,e^{-{p_\perp^2}/{M_{h}^2}}\,,
 \eea
and $-1\le N^{\T}_q\le 1$, $-1 \le N^{\C}_q \le 1$. We assume
$\avk \T = \avk$. The helicity distributions $\Delta q(x)$ are taken
from Ref.~\cite{Gluck:2000dy}. Notice that with these choices both
the transversity and the Collins function automatically obey their
proper positivity bounds.

Using these parameterizations we obtain the following expression for
$A^{\sin (\phi_h+\phi_S)}_{UT}$:
 \bea
&& \hspace*{-0.7cm}A^{\sin (\phi_h+\phi_S)}_{UT} = \frac{P_T}{M_{h}}
\frac{1-y}{1+(1-y)^2} \,C(P_T) \nonumber\\
&& \hspace*{-0.6cm}\times\frac{\sum_q e_q^2 \,
 {\cal N}^{\T}_q\!(x)
\left[f_{q/p}(x)+\Delta q(x) \right] {\cal N}^{\C}_q\!(z)
D_{h/q}(z)} { \sum_q e_q^2 \, f_{q/p}(x) D_{h/q}(z)}\,,\nonumber\\
 \label{sin-asym-final}
 \eea
where $C(P_T)$ is given by~\cite{Anselmino:2007fs}
 \be
C(P_T) = \sqrt{2e} \, \frac{\avp ^2 \C}{\avp} \,
\frac{e^{-P_T^2/\avPT \C}}{\avPT ^2 \C}
\frac{\avPT}{e^{-P_T^2/\avPT}}\,,
 \ee
with
 \bea
\avp\! \C= \frac{M_{h}^2 \avp}{M_{h}^2 +\avp}\,, &&\hspace*{-0.6cm}
 \avPT_{_{\!(C)}} =\avp_{_{\!(C)}} +z^2\avk \,.\nonumber\\
 \eea

When data or phenomenological information at different $Q^2$ values
are considered, we take into account, at leading order (LO), the QCD
evolution of the
transversity distribution. 
For the Collins FF, $\Delta ^N\! D_{h/\qup}$, as its scale
dependence is unknown, we tentatively assume the same $Q^2$
evolution as for the unpolarized FF, $D_{h/q}$.

By performing a best fit of the measurements of HERMES, COMPASS and
Belle Collaborations we then fix the free parameters, $\alpha,
\beta, \gamma, \delta, N^{\T}_q, N^{\C}_q$ and $M_h$ appearing in
$A^{\sin(\phi_h+\phi_S)}_{UT}$ ($q=u,d$).

\subsection{$e^+e^-\to h_1 h_2\, X$ processes}

One might think that hadron production in $e^+e^-$ collisions is
the cleanest process for the study of TMD polarized fragmentation
functions, like the Collins function, thanks to the lack of
corresponding TMD effects in the initial state. However, in the
process $e^+e^-\to q \, \bar{q}$ there is no transverse polarization
transfer to a single, on-shell final quark. Therefore, the single
Collins effect, i.e.~the asymmetry in the distribution around the
jet thrust axis (given by the fragmenting quark direction) of
hadrons produced in the quark fragmentation, cannot be measured.
Instead, in hadron production from $e^+e^-\to q\,\bar{q}\to 2\,{\rm
jets}$ events, the Collins effect can be observed when the quark and
the antiquark are considered \emph{simultaneously}. The Belle
Collaboration at the KEK-B asymmetric-energy $e^+e^-$ storage rings
have in fact performed a measurement of azimuthal hadron-hadron
correlations for inclusive charged dihadron production,
$e^+e^-\to\pi^+\pi^- X$~\cite{Abe:2005zx,Seidl:2008xc}. This
asymmetry has been interpreted as a direct measure of the Collins
effect, involving the convolution of two Collins functions.

Two methods have been adopted in the experimental analysis performed
by Belle. These can be schematically described as (for details
and definitions see, e.g., Refs.~\cite{Boer:1997mf,Anselmino:2007fs,Seidl:2008xc}): \\
$i)$ the ``$\cos(\varphi_1 + \varphi_2)$ method'' in the
Collins-Soper frame where the jet thrust axis is used as the $\hat
z$ direction and the $e^+e^-\to q \, \bar q$ scattering defines the
$\hat{xz}$ plane; \\
$ii)$ the ``$\cos(2\varphi_0)$ method'', using the Gottfried-Jackson
c.m.~frame where one of the produced hadrons ($h_2$) identifies the
$\hat z$ direction and the $\hat{xz}$ plane is determined by the
lepton and the $h_2$ directions. There will then be another relevant
plane, determined by $\hat z$ and the direction of the other
observed hadron $h_1$, at an angle $\varphi_0$ with respect to the
$\hat{xz}$ plane.

In both cases one integrates over the magnitude of the intrinsic
transverse momenta of the hadrons with respect to the fragmenting
quarks. For the $\cos(\varphi_1 + \varphi_2)$ method the
cross section for the process $e^+e^-\to h_1 h_2 \, X$ reads:
 \bea
&&\hspace*{-0.7cm}\frac{d\sigma ^{e^+e^-\to h_1 h_2 X}}
{dz_1\,dz_2\,d\cos\theta\,d(\varphi_1+\varphi_2)}\nonumber\\
&&\hspace*{-0.7cm} =\frac{3\alpha^2}{4s} \, \sum _q e_q^2 \, \Big\{
 (1+\cos^2\theta)\,D_{h_1/q}(z_1)\,D_{h_2/\bar q}(z_2)
\nonumber \\
&&\hspace*{-0.7cm}+\Big.\frac{\sin^2\theta}{4}\,\cos(\varphi_1\!+\!\varphi_2)\,
\Delta ^N\! D _{h_1/q^\ua}\!(z_1)\, \Delta ^N\! D _{h_2/\bar
q^\ua}\!(z_2)\Big\} \!,\nonumber\\
\label{int-Xs-belle}
 \eea
where $\theta$ is the angle between the lepton direction and the
thrust axis and
 \be
\Delta ^N\! D _{h/q^\ua}(z)\equiv \int d^2\bpp \Delta ^N\!
D_{h/q^\ua} (z,\pp)\,. \label{coll-mom}
 \ee

Normalizing to the azimuthal averaged unpolarized cross section one
has:
 \bea
 && A_{12}(z_1,z_2,\theta,\varphi_1 + \varphi_2)  \nonumber\\
&& \equiv \frac{1} {\langle d\sigma \rangle} \> \frac{d\sigma
^{e^+e^-\to h_1 h_2 X}}{dz_1\,dz_2\,d\cos\theta\, d(\varphi_1 +
\varphi_2)}
\nonumber\\
&&  =1+\frac{1}{4}\,\frac{\sin^2\theta}{1+\cos^2\theta}\,
\cos(\varphi_1+\varphi_2)\, \nonumber \\ && \times \frac{\sum_q
e^2_q \, \Delta ^N\! D_{h_1/q^\ua}(z_1)\,
 \Delta ^N\! D_{h_2/\bar q^\ua}(z_2)}{\sum_q e^2_q D _{h_1/q}(z_1)\,
 D _{h_2/\bar q}(z_2)}\,\cdot \label{A12g}
 \eea

{F}or the $\cos(2\varphi_0)$ method, where the Gaussian ansatz
(\ref{coll-funct}) becomes extremely helpful, the analogue of
Eq.~(\ref{A12g}) reads
 \bea
 && A_0(z_1,z_2,\theta_2,\varphi_0) \nonumber\\
&& = 1+\frac{1}{\pi}\,\frac{z_1\,z_2}{z_1^2+z_2^2}\,
\frac{\sin^2\theta_2}{1+\cos^2\theta_2}\,\cos(2\varphi_0)\,
\nonumber \\
&& \times \frac{\sum_q e^2_q \, \Delta ^N\! D _{h_1/q^\ua}(z_1)\,
 \Delta ^N\! D  _{h_2/\bar q^\ua}(z_2)}{\sum_q e^2_q D _{h_1/q}(z_1)\,
 D _{h_2/\bar q}(z_2)}\,, \label{A0g}
 \eea
where $\theta_2$ is now the angle between the lepton and the $h_2$
hadron directions.

To eliminate false asymmetries, the Belle
Collaboration~\cite{Seidl:2008xc} consider the ratio of unlike-sign
to like-sign pion pair production, $A_U$ and $A_L$.

For fitting purposes, it is usually convenient to express these
relations in terms of favoured and unfavoured fragmentation
functions,
 \bea && D_{\pi^+/u,\bar d} = D_{\pi^-/d,\bar u} \equiv D_{\rm fav} \label{fav} \,,\\
&& D_{\pi^+/d,\bar u} = D_{\pi^-/u,\bar d} = D_{\pi^\pm/s,\bar s}
\equiv D_{\rm unf}, \label{unf}
 \eea
and similarly for the $\Delta^N\! D$'s.

\section{Results}

A combined fit of SIDIS asymmetries together with $e^+ e^- \to h_1
h_2 X$ data, Eqs.~(\ref{sin-asym-final},\ref{A12g},\ref{A0g}),
allows the simultaneous extraction of the transversity distribution
and the Collins fragmentation functions. We assume flavour
independent values of $\alpha$ and $\beta$ (neglecting transversity
distributions of sea quarks) and, similarly, we assume that $\gamma$
and $\delta$ are the same for favoured and unfavoured Collins
fragmentation functions; we then remain with a total number of 9
parameters.

The first study along this line was presented in
Ref.~\cite{Anselmino:2007fs}. Here we repeat the analysis,
exploiting the new high-precision data recently released by the
HERMES~\cite{Diefenthaler:2007rj} and COMPASS~\cite{:2008dn}
Collaborations for SIDIS, and by the Belle
Collaboration~\cite{Seidl:2008xc} for $e^+e^-$ annihilation
processes, in order to refine and reduce the uncertainty of the
previous extraction.

\begin{figure}[h!]
\begin{center}
\includegraphics[width=0.35\textwidth,bb= 10 140 550 660,angle=-90]
{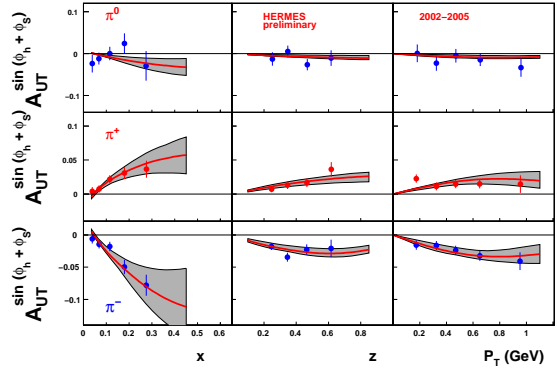} 
\end{center}
\vskip -0.5cm \caption{\label{fig:hermes} Fit of
HERMES~\cite{Diefenthaler:2007rj} data. The shaded area corresponds
to the statistical uncertainty in the parameter values, see text.}
\end{figure}

\begin{figure}[h!t]
\begin{center}
\includegraphics[width=0.35\textwidth,bb= 10 140 550 660,angle=-90]
{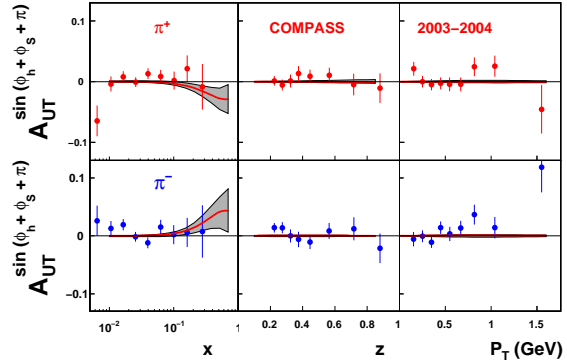}
\end{center}
\vskip -0.5cm \caption{\label{fig:compass} Fit of
COMPASS~\cite{:2008dn} data. The shaded area corresponds to the
statistical uncertainty in the parameter values, see text. The extra
$\pi$ phase in addition to $\phi_h+\phi_S$ comes from the different
convention adopted by COMPASS.}
\end{figure}

New data from COMPASS operating on a transversely polarized hydrogen
target have recently been released~\cite{Levorato:2008tv}: these are
not included in the fit but compared with our predictions.

The two sets of Belle data, coming from two analyses of the same
experimental events, are not independent. Therefore we include only
one set of data in the fit, either $A_{0}$ or $A_{12}$ data. In this
analysis we report the results obtained by using $A_{12}$ data, the
$\cos(\varphi_1 + \varphi_2)$ method. The consequences of fitting
$A_{0}$ instead of $A_{12}$ are presently under investigation.

\begin{figure}[h!t]
\begin{center}
\includegraphics[width=0.35\textwidth,bb= 80 40 500 660,angle=-90]
{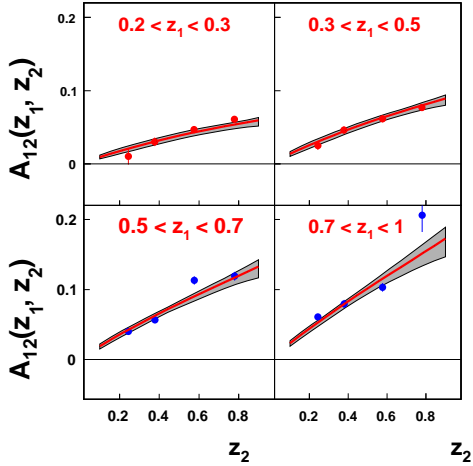} \hspace*{0.3cm}
\end{center}
\caption{\label{fig:belle} Fit of the Belle~\cite{Seidl:2008xc} data
on the $A_{12}$ asymmetry (the $\cos (\varphi_1 + \varphi_2)$
method). }
\end{figure}

\begin{figure}[h!b]
\begin{center}
\includegraphics[width=0.35\textwidth,bb= 80 40 500 660,angle=-90]
{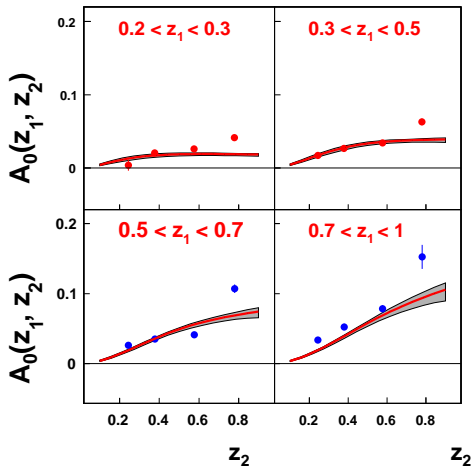}
\end{center}
\caption{\label{fig:belle2} Comparison of our predictions with
Belle~\cite{Seidl:2008xc} data for the $A_{0}$ Belle asymmetry (the
$\cos(2\,\varphi_0)$ method). }
\end{figure}

In Figs.~\ref{fig:hermes} and \ref{fig:compass} we show the best fit
to the HERMES~\cite{Diefenthaler:2007rj} and COMPASS~\cite{:2008dn}
data, respectively. Notice that the $\pi^0$ data (HERMES) have not
been used in the fit; in Fig.~\ref{fig:hermes} we show our
estimates, based on the extracted transversity and Collins
functions, and compare them to data. Fig.~\ref{fig:belle} shows the
fit to the Belle $A_{12}$ asymmetry, whereas in
Fig.~\ref{fig:belle2} our predictions for the $A_{0}$ asymmetry are
compared with data~\cite{Seidl:2008xc}.

The curves shown are evaluated using the central values of the
parameters in Table~\ref{fitpar}, while the shaded areas correspond
to a two-sigma deviation at 95.45\% Confidence Level (for details
see Appendix A of Ref.~\cite{Anselmino:2008sga}).

\begin{table}[h!]
\caption{Best values of the free parameters for the $u$ and $d$
transversity distribution functions and for the favoured and
unfavoured Collins fragmentation functions. We obtain $\chi^2/{\rm
d.o.f.} = 1.3$.  Notice that the errors generated by MINUIT are
strongly correlated, and should not be taken at face value. The
significant fluctuations in our results are shown by the shaded
areas in the plots.\label{fitpar}}
\renewcommand{\tabcolsep}{0.4pc} 
\renewcommand{\arraystretch}{1.2} 
\begin{tabular}{@{}ll}
 \hline
 $N_{u}^T$ =  $0.64 \pm 0.34$ & $N_{d}^T$ =  $ -1.00  \pm
0.02$      \\
$\alpha$ =  $0.73  \pm  0.51$ & $\beta$  = $0.84  \pm  2.30$ \\
 \hline
 $N_{fav}^C$  = $0.44  \pm 0.07$ & $N_{unf}^C$   = $-1.00 \pm 0.06$ \\
 $\gamma$  = $0.96  \pm  0.08$  & $\delta$   = $0.01  \pm  0.05$    \\
$M^2_h = 0.91 \pm 0.52$ GeV$^2$ &\\
\hline
\end{tabular}
\end{table}

\begin{figure}[t]
\begin{center}
\includegraphics[width=0.35\textwidth,bb= 10 120 500 660,angle=-90]
{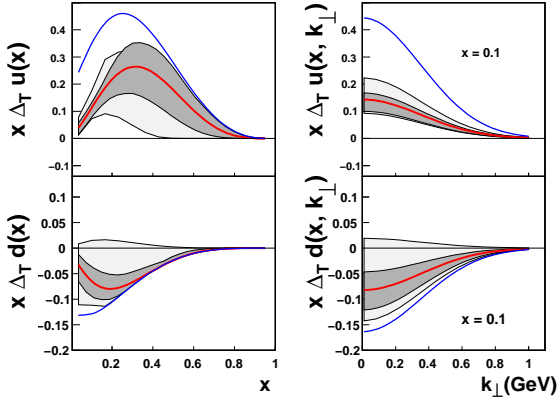}\hskip 2cm
\end{center}
\caption{\label{fig:transv} The transversity distribution functions
for $u$ and $d$ flavours as determined by our global fit, at $Q^2$ =
2.4 GeV$^2$; we also show the Soffer bound~\cite{Soffer:1994ww}
(highest or lowest lines) and the (wider) uncertainty bands of our
previous extraction~\cite{Anselmino:2007fs}.}
\end{figure}

\begin{figure}[h!t]
\begin{center}
\includegraphics[width=0.35\textwidth,bb= 10 130 500 660,angle=-90]
{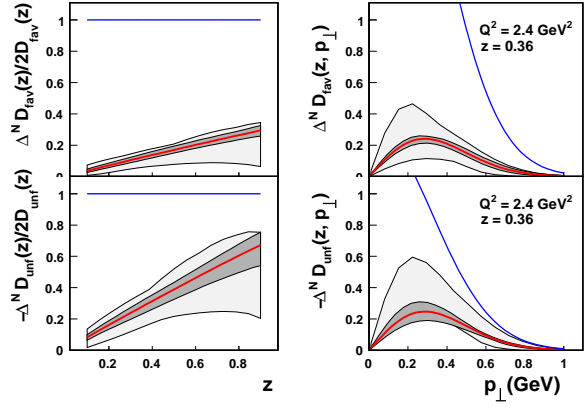}
\end{center}
\caption{\label{fig:coll} Favoured and unfavoured Collins
fragmentation functions as determined by our global fit, at $Q^2$ =
2.4 GeV$^2$; we also show the positivity bound and the (wider)
uncertainty bands as obtained in Ref.~\cite{Anselmino:2007fs}.}
\end{figure}

Table~\ref{fitpar} collects the results of our best fit to the new
data sets~\cite{Diefenthaler:2007rj,:2008dn,Seidl:2008xc}, while in
Figs.~\ref{fig:transv} and \ref{fig:coll} we show our updated
transversity distribution and Collins fragmentation functions
together with the uncertainty bands of our previous
extraction~\cite{Anselmino:2007fs}. We can definitely say that the
two extractions are compatible with each other, with the new error
bands strongly reduced. The transversity for up quarks results now
larger (compared to our previous extraction), while that for down
quarks is better constrained in sign and non compatible with zero.
In this respect the new data from SIDIS have been crucial. It is
worth noticing that while the transversity for up quarks is strongly
constrained by HERMES data, in particular through the positive pion
azimuthal asymmetry, the addition of COMPASS deuteron data to the
fit allows a better determination of $\Delta_T d$. We recall here
that, in analyzing SIDIS data, we have assumed the transversity
distributions  for sea quarks and antiquarks to vanish. The
extracted Collins FFs are well constrained and much smaller than
their positivity bounds, with the unfavoured Collins function large
in size and negative, consistently with other
extractions~\cite{Efremov:2006qm,Vogelsang:2005cs,Anselmino:2007fs}.

A word of caution has to be added here since SIDIS data (HERMES and
COMPASS) are collected at a much smaller scale ($Q^2\simeq 2.5$
GeV$^2$) compared to the Belle data ($Q^2=110$ GeV$^2$).

Both azimuthal asymmetries in SIDIS and in $e^+e^-$ collisions
involve spin and TMD functions whose behaviour upon scale variation
should be described in the context of Collins-Soper
factorization~\cite{Collins:1981uk,Ji:2004wu}. Beyond tree level
this would result in a soft factor entering TMD convolutions, with
the corresponding Sudakov suppression. This, as discussed in
Refs.~\cite{Boer:2008fr,Boer:2008mz}, might imply an underestimation
of the Collins function as extracted at tree level from the
azimuthal asymmetry at Belle. Hence the combined extraction of the
transversity from SIDIS at a lower $Q^2$ (less Sudakov suppression),
might lead to an overestimation of $\Delta_T q$. This issue is
currently under study. Here, as in Ref.~\cite{Anselmino:2007fs}, the
$Q^2$ dependence of the Collins FF {\it is} included assuming it to
be the same as that of the unpolarized fragmentation function,
$D_{h/q}$: although this might not be the proper evolution, it
should mitigate the above-mentioned effect.

\begin{figure}[h!t]
\begin{center}
\includegraphics[width=0.35\textwidth,bb= 10 120 500 660,angle=-90]
{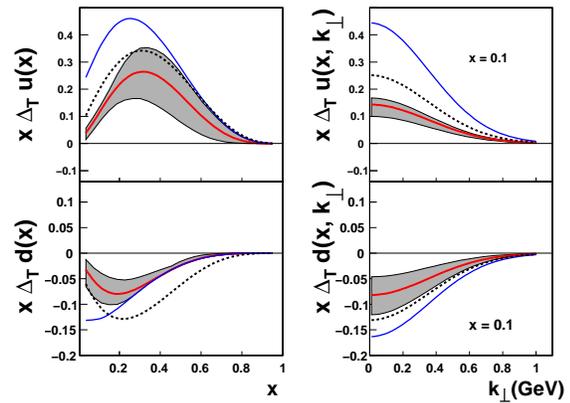}
\end{center}
\caption{\label{fig:transv:helicity}  Comparison of the extracted
transversity (solid line) with the helicity distribution (dashed
line) at $Q^2=2.4$ GeV$^2$. The Soffer bound~\cite{Soffer:1994ww}
(blue solid line) is also shown.}
\end{figure}

As it is well known, in a non relativistic theory the helicity and
the transversity distributions should be equal. We then show in
Fig.~\ref{fig:transv:helicity} the
extracted transversity distribution together with the helicity
distribution of Ref.~\cite{Gluck:2000dy} at $Q^2=2.4$ GeV$^2$. It results
that, both for $u$ and $d$ quarks, we have $|\Delta_T q| < |\Delta q|$.

\begin{figure}[t]
\begin{center}
\includegraphics[width=0.35\textwidth,bb= 60 110 560 660,angle=0]
{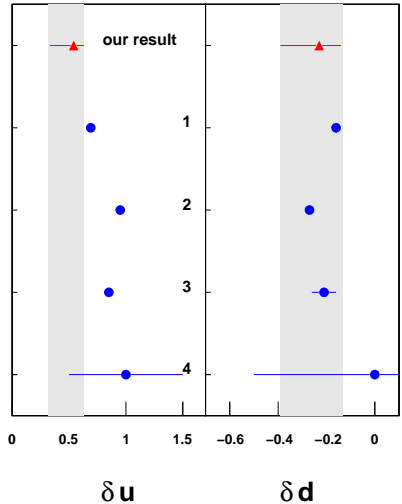}
\end{center}
\caption{\label{fig:tensor}  Tensor charge from different models
compared to our result. 1: Quark-diquark model of
Ref.~\cite{Cloet:2007em}, 2: Chiral quark soliton model of
Ref.~\cite{Wakamatsu:2007nc}, 3: Lattice QCD~\cite{Gockeler:2005cj},
4: QCD sum rules~\cite{He:1994gz}.}
\vspace{-1pc}
\end{figure}

Another interesting quantity, related to the first $x$-moment of
the transversity distribution, is the tensor charge:
 \be
 \delta q = \int_0^1\! dx \, (\Delta_T q - \Delta_T \bar
q) =\int_0^1\!dx \, \Delta_Tq
 \ee
where the last equality is valid for zero antiquark transversity, as
assumed in our approach. From our analysis we get, at $Q^2 = 0.8$
GeV$^2$,
 \be
 \delta u = 0.54^{+0.09}_{-0.22} \hspace*{1cm}
 \delta d =-0.23^{+0.09}_{-0.16}\,.
 \ee
Such values are quite close to various model
predictions~\cite{Cloet:2007em,Wakamatsu:2007nc,Gockeler:2005cj,He:1994gz}
for tensor charges which span the ranges $0.5 \le \delta u \le 1.5$
and $-0.5 \le \delta d \le 0.5$ (see Fig.~\ref{fig:tensor}). In this
context it is worth mentioning a subtle point concerning the strong
scale dependence of the tensor charge, recently addressed in
Ref.~\cite{Wakamatsu:2008ki}.  For the effective models of baryons,
as those referred to above, the choice of their starting energy
scale and their $Q^2$ evolution could play a significant role and,
eventually, mask the true nature of the model. Consequently, the
results shown in Fig.~\ref{fig:tensor}, where our LO
phenomenological extraction seems in better agreement with the
quark-diquark model of Ref.~\cite{Cloet:2007em} than with other
models, should be taken with some care. A safer quantity, totally
scale independent, and therefore easy to compare with, would be the
ratio of two tensor charges. From our fit, for instance, we obtain
$\delta d/\delta u = -0.42^{+0.0003}_{-0.20}$, and all model
predictions considered above would fall within our uncertainty band,
as shown in Fig.~7 of Ref.~\cite{Wakamatsu:2008ki}.

\section{Predictions}

\begin{figure}[t]
\begin{center}
\includegraphics[width=0.35\textwidth,bb= 20 100 560 660,angle=-90]
{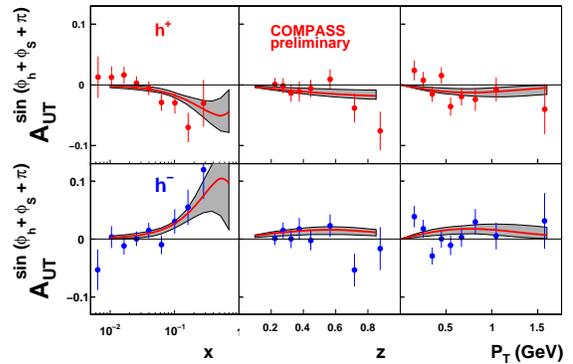}
\end{center}
\vskip -0.5cm\caption{\label{fig:compass-proton}Predictions for the
single spin asymmetry $A_{UT}^{\sin(\phi_h+\phi_S+\pi)}$ compared to
preliminary data by the COMPASS experiment operating with a
transversely polarized hydrogen target~\cite{Levorato:2008tv}.}
\end{figure}

We now use the extracted transversity and Collins functions to give
predictions for new measurements performed or planned at COMPASS and
JLab. The transverse single spin asymmetry
$A_{UT}^{\sin(\phi_h+\phi_S)}$ has been recently measured by the
COMPASS experiment operating with a polarized hydrogen target
(rather than a deuterium one). In Fig.~\ref{fig:compass-proton} we
show our predictions compared with these preliminary data. The
agreement is excellent.

\begin{figure}[h!t]
\begin{center}
\includegraphics[width=0.35\textwidth,bb= 20 100 560 660,angle=-90]
{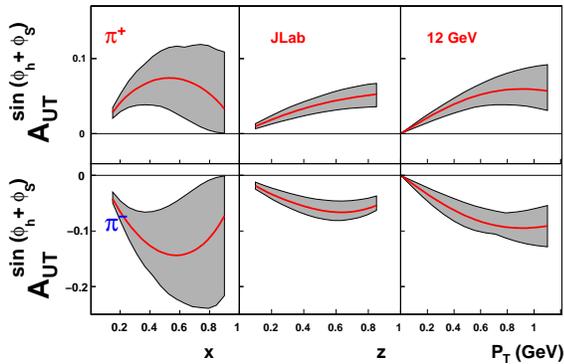}
\end{center}
\vskip -0.5cm \caption{\label{fig:jlab} Estimates of the single spin
asymmetry $A_{UT}^{\sin(\phi_h+\phi_S)}$ for JLab operating with
proton target.}
\end{figure}

In Fig.~\ref{fig:jlab} we present our estimates for JLab operating
with a proton target at 12 GeV. Notice that JLab results will give
important information on the large $x$ region, which is left
basically unconstrained by the present SIDIS data from HERMES and
COMPASS. In this region our estimates must be taken with some care.
We recall that the large $x$ behaviour of our parameterization is
controlled by the same $\beta$ parameter for $\Delta_T u$ and
$\Delta_T d$ (since present data do not cover the large $x$ region).
The same is true for the Collins fragmentation functions, whose
large $z$ behaviour is driven by the same parameter $\delta$ for
favoured and unfavoured Collins FFs. On the other hand for the small
to medium $x$ region, well constrained by SIDIS measurements, data
support the choice of a universal behaviour $x^{\alpha}$ for
$\Delta_T u$ and $\Delta_T d$. The future JLab measurements, which
will extend to larger $x$ values, will test the validity of this
approximations.

\section{Conclusions}

We have performed a re-analysis of recent high-precision
experimental data on spin azimuthal asymmetries which involve the
transversity distributions of $u$ and $d$ quarks and the Collins
fragmentation functions. The values of the 9 free parameters are
fixed by simultaneously best fitting the HERMES, COMPASS and Belle
data.

All data can be accurately described, leading to the extraction of
the favoured and unfavoured Collins functions, in agreement with
similar results previously obtained in the
literature~\cite{Vogelsang:2005cs,Efremov:2006qm,Anselmino:2007fs}.
In addition, we have improved the extraction of the so far poorly
known transversity distributions for $u$ and $d$ quarks, $\Delta_T
u(x)$ and $\Delta_T d(x)$. They turn out to be opposite in sign,
with $|\Delta_T d(x)|$ smaller than $|\Delta_T u(x)|$, and both
smaller than their Soffer bound~\cite{Soffer:1994ww}. The previous
uncertainty bands are strongly reduced by the present analysis. The
new distributions are compatible with our previous
extraction~\cite{Anselmino:2007fs} and close to some model
predictions for the transversity distribution.

The extracted transversity distributions and the Collins
fragmentation functions allow to compute the azimuthal asymmetry
$A_{UT}^{\sin(\phi_h+\phi_S)}$ for any SIDIS process. In particular,
our predictions for the COMPASS measurements with a proton target
are in very good agreement with preliminary data, while the large
$x$ behaviour of $\Delta_T q$, yet unknown, could be explored by
JLab experiments. These will provide further important tests of our
complete understanding of the partonic properties which are at the
origin of SSAs.

Further expected data from Belle will allow to study in detail not
only the $z$ dependence of the Collins functions, but also their
$p_\perp$ dependence.

The TMD approach to azimuthal asymmetries in SIDIS and $e^+e^- \to
h_1 h_2 \, X$ processes has definitely opened a new powerful way of
studying the nucleon structure and fundamental QCD properties.

\section*{Acknowledgments}
U.D.~would like to thank the organizers of this very interesting
workshop for their kind hospitality. We acknowledge the support of
the European Community - Research Infrastructure Activity under the
FP6 ``Structuring the European Research Area'' program
(HadronPhysics, RII3-CT-2004-506078). M.A.,~M.B., and
A.P.~acknowledge partial support by MIUR under PRIN 2006 and by the
Helmholtz Association through funds provided to the virtual
institute ``Spin and strong QCD''(VH-VI-231).


\begin{thebibliography}{10}

\bibitem{Ralston:1979ys}
J.P. Ralston and D.E. Soper,
\newblock Nucl. Phys. B152 (1979) 109.

\bibitem{Barone:2001sp}
V. Barone, A. Drago and P.G. Ratcliffe,
\newblock Phys. Rep. 359 (2002) 1, hep-ph/0104283.

\bibitem{Martin:1999mg}
O. Martin et~al.,
\newblock Phys. Rev. D60 (1999) 117502, hep-ph/9902250.

\bibitem{Barone:2005pu}
PAX, V. Barone et~al.,
\newblock (2005), hep-ex/0505054.

\bibitem{Anselmino:2004ki}
M. Anselmino et~al.,
\newblock Phys. Lett. B594 (2004) 97, hep-ph/0403114.

\bibitem{Efremov:2004qs}
A.V. Efremov, K. Goeke and P. Schweitzer,
\newblock Eur. Phys. J. C35 (2004) 207, hep-ph/0403124.

\bibitem{Pasquini:2006iv}
B. Pasquini, M. Pincetti and S. Boffi,
\newblock Phys. Rev. D76 (2007) 034020, hep-ph/0612094.

\bibitem{Mukherjee:2003pf}
A. Mukherjee, M. Stratmann and W. Vogelsang,
\newblock Phys. Rev. D67 (2003) 114006, hep-ph/0303226.

\bibitem{Mukherjee:2005rw}
A. Mukherjee, M. Stratmann and W. Vogelsang,
\newblock Phys. Rev. D72 (2005) 034011, hep-ph/0506315.

\bibitem{Bacchetta:2008wb}
A. Bacchetta et~al.,
\newblock (2008), arXiv:0812.0611 [hep-ph].

\bibitem{Efremov:1992pe}
A.V. Efremov, L. Mankiewicz and N.A. Tornqvist,
\newblock Phys. Lett. B284 (1992) 394.

\bibitem{Collins:1994ax}
J.C. Collins and G.A. Ladinsky,
\newblock (1994), hep-ph/9411444.

\bibitem{Artru:1995zu}
X. Artru and J.C. Collins,
\newblock Z. Phys. C69 (1996) 277, hep-ph/9504220.

\bibitem{Jaffe:1997hf}
R.L. Jaffe, X.m. Jin and J. Tang,
\newblock Phys. Rev. Lett. 80 (1998) 1166, hep-ph/9709322.

\bibitem{Bacchetta:2003vn}
A. Bacchetta and M. Radici,
\newblock Phys. Rev. D69 (2004) 074026, hep-ph/0311173.

\bibitem{Collins:1992kk}
J.C. Collins,
\newblock Nucl. Phys. B396 (1993) 161, hep-ph/9208213.

\bibitem{Abe:2005zx}
Belle, K. Abe et~al.,
\newblock Phys. Rev. Lett. 96 (2006) 232002, hep-ex/0507063.

\bibitem{Airapetian:2004tw}
HERMES, A. Airapetian et~al.,
\newblock Phys. Rev. Lett. 94 (2005) 012002, hep-ex/0408013.

\bibitem{Ageev:2006da}
COMPASS, E.S. Ageev et~al.,
\newblock Nucl. Phys. B765 (2007) 31, hep-ex/0610068.

\bibitem{Anselmino:2007fs}
M. Anselmino et~al.,
\newblock Phys. Rev. D75 (2007) 054032, hep-ph/0701006.

\bibitem{Diefenthaler:2007rj}
HERMES, M. Diefenthaler,
\newblock (2007), arXiv:0706.2242 [hep-ex].

\bibitem{:2008dn}
COMPASS, M. Alekseev et~al.,
\newblock (2008), arXiv:0802.2160 [hep-ex].

\bibitem{Seidl:2008xc}
Belle, R. Seidl et~al.,
\newblock Phys. Rev. D78 (2008) 032011, arXiv:0805.2975 [hep-ex].

\bibitem{Mulders:1995dh}
P. Mulders and R. Tangerman,
\newblock Nucl. Phys. B461 (1996) 197, hep-ph/9510301.

\bibitem{Boer:1997nt}
D. Boer and P.J. Mulders,
\newblock Phys. Rev. D57 (1998) 5780, hep-ph/9711485.

\bibitem{Bacchetta:2006tn}
A. Bacchetta et~al.,
\newblock JHEP 02 (2007) 093, hep-ph/0611265.

\bibitem{D'Alesio2007jt}
U. D'Alesio and F. Murgia,
\newblock Prog. Part. Nucl. Phys. 61 (2008) 394, arXiv:0712.4328 [hep-ph].

\bibitem{Collins:1981uk}
J.C. Collins and D.E. Soper,
\newblock Nucl. Phys. B193 (1981) 381.

\bibitem{Ji:2004xq}
X.-d. Ji, J.-p. Ma and F. Yuan,
\newblock Phys. Lett. B597 (2004) 299, hep-ph/0405085.

\bibitem{Ji:2004wu}
X.-d. Ji, J.-p. Ma and F. Yuan,
\newblock Phys. Rev. D71 (2005) 034005, hep-ph/0404183.

\bibitem{Metz:2002iz}
A. Metz,
\newblock Phys. Lett. B549 (2002) 139.

\bibitem{Collins:2004nx}
J.C. Collins and A. Metz,
\newblock Phys. Rev. Lett. 93 (2004) 252001, hep-ph/0408249.

\bibitem{Boer:2001he}
D. Boer,
\newblock Nucl. Phys. B603 (2001) 195, hep-ph/0102071.

\bibitem{Bacchetta:2004jz}
A. Bacchetta et~al.,
\newblock Phys. Rev. D70 (2004) 117504, hep-ph/0410050.

\bibitem{Anselmino:2005nn}
M. Anselmino et~al.,
\newblock Phys. Rev. D71 (2005) 074006, hep-ph/0501196.

\bibitem{Gluck:1998xa}
M. Gl{\"u}ck, E. Reya and A. Vogt,
\newblock Eur. Phys. J. C5 (1998) 461, hep-ph/9806404.

\bibitem{deFlorian:2007aj}
D. de~Florian, R. Sassot and M. Stratmann,
\newblock Phys. Rev. D75 (2007) 114010, hep-ph/0703242.

\bibitem{Gluck:2000dy}
M. Gl{\"u}ck et~al.,
\newblock Phys. Rev. D63 (2001) 094005, hep-ph/0011215.

\bibitem{Boer:1997mf}
D. Boer, R. Jakob and P.J. Mulders,
\newblock Nucl. Phys. B504 (1997) 345, hep-ph/9702281.

\bibitem{Levorato:2008tv}
COMPASS, S. Levorato,
\newblock (2008), arXiv:0808.0086 [hep-ex].

\bibitem{Anselmino:2008sga}
M. Anselmino et~al.,
\newblock (2008), arXiv:0805.2677 [hep-ph].

\bibitem{Efremov:2006qm}
A.V. Efremov, K. Goeke and P. Schweitzer,
\newblock Phys. Rev. D73 (2006) 094025, hep-ph/0603054.

\bibitem{Vogelsang:2005cs}
W. Vogelsang and F. Yuan,
\newblock Phys. Rev. D72 (2005) 054028, hep-ph/0507266.

\bibitem{Boer:2008fr}
D. Boer,
\newblock Nucl. Phys. B806 (2009) 23, arXiv:0804.2408 [hep-ph].

\bibitem{Boer:2008mz}
D. Boer,
\newblock (2008), arXiv:0808.2886 [hep-ph].

\bibitem{Soffer:1994ww}
J. Soffer,
\newblock Phys. Rev. Lett. 74 (1995) 1292, hep-ph/9409254.

\bibitem{Cloet:2007em}
I.C. Cloet, W. Bentz and A.W. Thomas,
\newblock Phys. Lett. B659 (2008) 214, arXiv:0708.3246 [hep-ph].

\bibitem{Wakamatsu:2007nc}
M. Wakamatsu,
\newblock Phys. Lett. B653 (2007) 398, arXiv:0705.2917 [hep-ph].

\bibitem{Gockeler:2005cj}
QCDSF, M. Gockeler et~al.,
\newblock Phys. Lett. B627 (2005) 113, hep-lat/0507001.

\bibitem{He:1994gz}
H.-x. He and X.-d. Ji,
\newblock Phys. Rev. D52 (1995) 2960, hep-ph/9412235.

\bibitem{Wakamatsu:2008ki}
M. Wakamatsu,
\newblock (2008), arXiv:0811.4196 [hep-ph].

\end{thebibliography}
\end{document}